# A Survey of Event-triggered Control for Nonlinear Multiagent Systems with Guaranteed Steady-State Performance

Gurmu Meseret Debele

Email: mersimoidebele2027@gmail.com

School of Automation Central South University, Hunan, China.

**Abstract**—With the gradual advancement of a novel idea of the distributed control of the multiagent systems, an event-triggered control protocol has received significant research attention, especially in designing the controller for the nonlinear multiagent system. Compared to other widely used control conditions, the event-triggered control of the nonlinear system has a significant capability to improve resource utilization in real-life scenarios such as using and controlling the intelligent control input of each agent. It is worth mentioning that a group of interconnected agents have a network communication topology to transmit the feedback information state across the networked link. The transmission of information among a group of agents ensures that each agent reaches the consensus agreement cooperatively. The cooperative protocol of the distributed control of nonlinear multiagent system also ensures the proper information flow between each agent, irrespective of communication delays, variability of environment, and switching of the communication topology via the event-triggered control protocol. Consequently, event-triggered control for nonlinear multiagent systems via steady-state performance will be investigated in this paper. The steady-state performances of a nonlinear closed-loop system demonstrate the stabilization, output regulation, and output synchronization problem of the nonlinear system using proper control protocol to achieve a consensus in a multiagent system will also be discussed. Based on the steady-state conditions of the nonlinear system, the consensus agreement among the agents will be realized.

# 1 Introduction

The rapid developments of computer technologies, such as the emergence of embedded systems, have been seen for the past couple of decades, which results in the advancement of discrete-time to its highest stage. The advantages of this analog equipment are that they are small, flexible, decrease energy consumption, and save the installation cost. So, they are becoming parts of a wide area of applications. In the case of discrete-time systems, even if selecting an appropriate sampling interval is challenging, only sampling signals at the discrete-time instant are used. In terms of some small but highly integrated embedded systems, the energy supply could not be inexhaustible because of the limited space of the power module.

Traditionally, the system tunes the state of its actuator at every sampling instant using the time-driven control scheme. Such time-triggered control mechanisms may cause a significant frequent change to the state of the actuators, which could lead to unnecessary energy consumption or even actuator damage. Therefore, various researchers have studied an event-triggered control mechanism to overcome these disadvantages. That means, in event-triggered control mechanisms, the agents are modulated after the system meets certain conditions given in the control system. In other words, the control protocol for each agent will be changed after the state of each agent approach to a specific measurement value or condition; in that case, it will be guaranteed that the stable system performance and efficiency of energy utilization. Let us consider an event-triggered control problem, in which we desire to maintain an agent state in a specific interval of a state as time goes to infinity. That is, the agent's control signals will be triggered only when the state leaves the predetermined interval; otherwise, the agent dynamics (i.e., state and controller) remain unchanged. When the agent state approaches its desired equilibrium state, the execution time instants of each control protocol of all agents become too close to each other; this phenomenon is called Zeno-behavior, and in this case, which could lead the system to infinite triggering in an instant of time [1]. Therefore, the event-triggered control strategies could be applicable by the existence of the lower bound of the execution's instants.

Despite its superior advantages over the time-driven control approach, the event-triggered control mechanisms impose some difficulties in the control or estimation synthesis, such as abandoning many indispensable executions to overcome the continuous triggering and meet the reasonable resource allocations. So, thorough care must be taken to balance the system performance and the resource allocations while using the event-triggered control scheme.

The distributed control of nonlinear multiagent systems has gained much research attention from different fields of study for its huge application. For example, in sensor fusion network, scheduling airlines, automated scheduling of highway systems, satellite clustering, cooperation control of mobile robots or multiple vehicles, formation control, and distributed estimation; and the consensus control of multiagent system has also received significant attention from many scientific researchers and experts such as the agent dynamics [2-3], network topology [4], rate of convergence [5], and the information transmission capability [6]. So, increasing the number of agents will arouse the agents' demands of reduced computational time and bandwidth.

Basically, an advanced control design may equip a small embedded processor installed on each agent to collect information from the neighboring nodes and update or trigger the controller using predefined control law. In this case, the controller update will be done with an event-triggered control protocol. Since the goal is to allow more agents with reduced computational cost and proper resource allocation, then an event-triggered control scheme is suitable. In [7-8], the stochastic event-driven mechanisms have been illustrated, and similarly, deterministic event-triggered feedback control has been found in [9].

Based on the above discussion, a deterministic event-triggered control scheme has been used for the large class of the cooperative control algorithms [11], and in [10-11], the distributed control design of multiagent system initiates the agent to trigger its control law whenever the controller of the neighbors is updated, and the state of each agent meet the required threshold value. The distributed event-triggered control in which the agent's controller does not have to update when the neighbor's control law update was provided by [12]. In [13], the event-triggered control mechanism is proposed to track the control problem of the second-order multiagent system and investigate the characteristics of the nonlinear system via distributed sliding mode control approach; and the event-triggered control is utilized to save the network communication resources by decreasing the sampling frequency of the system.

Recently, an event-triggered control-based feasible consensus algorithm has also been developed in [14]. The author has investigated the event-triggered control for an average-consensus problem of a single integrator system under undirected topology conditions, and the proposed consensus scheme allows each agent to monitor the state of its neighbor continuously. Moreover, the event-triggered control of the consensus problem of the double-integrator system has been presented in [15]. Despite the continuous monitoring of neighbor agents' state and the global network topology information, designing the triggering conditions is still needed. Therefore, in order to overcome this limitation, the fundamental triggering condition is that the designed triggering strategy is used to determine the changing rates of agent states is implemented in [16].

Motivated by the above conditions, [14] has presented the event-triggered control protocol for the average consensus of first and second-order Lipschitz under the undirected communication networks to develop the centralized and distributed event-triggered strategies. The event-triggered control system to achieve the output regulation problem and control the triggering frequency for each agent in a limited time in the system and the Zeno-behavior is excluded. [17] adopted an event-triggered control protocol to reduce the network communications among the neighboring agent, and the output regulation problem based on the assumption that all the states of the agents are measurables have been investigated in [18].

In [19], the consensus of the first-order discrete-time nonlinear multiagent system with time-varying communication topology was discussed. In this respect, an event-triggered control mechanism is employed to trigger or update the control input of each agent, and the sufficient condition of the consensus and the feasibility of the linear matrix inequality has been discussed by applying the Lyapunov function method. The communication strategy for the second-order multiagent system with nonlinear dynamics has been referenced in [20] to address the problem of the limitation of the communication channel resources and the event-triggered control mechanism; it is applied to get rid of the continuous usage of the sampling signals among the agents in the leader-follower multiagent systems. The event-activated control approach of the consensus problem of the distributed multiagent systems consisting of the single integrator dynamics are developed based on the control protocol and the local information only, and the event condition of the system is obtained by carrying out the convergence analysis of using the common Lyapunov function method.

## 2. The common steady-state condition of nonlinear system
### 2.1. Stabilization

When discussing the stability of a closed-loop system, it also includes a certain level of performance that the system could achieve in a given condition or design parameters. The primary purpose of designing a controller in a control design is eventually to achieve a certain degree of the system performance of the closed-loop system. This closed-loop system can be classified into a time-varying and time-invariant system. A Time-varying system is a system that depends on an external time-varying signal, while the time-invariant system does not change with a course of changing time. The stability of the nonlinear closed-loop system is all about its equilibrium

point(origin). It is also categorized into stability in whole, asymptotically stable, and globally asymptotically stable to support the idea of stable system performances. The steady-state performances of a high-level networked nonlinear system are based on the stabilization, regulation, and synchronization problems.

A nonlinear closed-loop system with equilibrium points at the origin, zero control input, and takes certain external disturbances; the system is said to achieve stabilization if and only if the system at the equilibrium point is globally asymptotically stable. The state of the network of the closed-loop system starts from its initial state in the convergence region of the system; as time goes by, the state of the system gradually converges to the equilibrium point or the origin point. In that case, the stabilization of the nonlinear system is easily realized. [21] proposed an event-triggered linear feedback control to study a global stabilization of the neutrally stable linear closed-loop system using an actuator saturation. When evaluating the state of the system, control input, and the history of an event-triggered control, the states of the system have been able to converge to an equilibrium state, guaranteeing that as time goes by, the global asymptotic stabilization is achieved.

The consensus problem of an event-triggered control for a high-order multiagent system under switching interaction topology is studied in [22]. The author proposed a novel model portioning approach where the high-order error is partitioned into stable and unstable subsystems by applying coordinate transformations. The unstable subsystem was analyzed by arbitrary switching topologies, and an appropriate controller will be designed to ensure its stability and guarantee that convergence of the error state of the system is within the interval of time. An error transformation-based controller design of a multiagent system in the non-affine form is presented by [23], and the radial basis function neural network (RBFNNs) together with the dynamic surface technique is used to approximate unknown nonlinear function and obtain the parameters to design the virtual controller via the first-order filter. Then an event-triggered control protocol is employed to reduce the network communication frequency and ensures that semi-globally uniformly bounded condition is satisfied to achieve the consensus tracking errors of an agent.

It is also known that consensus is a substantial dynamic behavior of the multiagent system under various communication topologies of each agent. It helps to understand the cooperation and coordination of multiagent systems [24], and consensus of the nonlinear multiagent systems will be achieved in many terminologies of control protocol such as leader-following, leaderless, decentralized, and formation control, and distributed control. Even though the consensus was achieved, [25] addressed the problem of the leader-following consensus of a multiagent system that the state of each agent will be updated based on the feedback information of the relative adjacent agents to reduce the error tracking between the leader and followers. In [26], the leader following event-triggered consensus problem of the multiagent system is proposed, but the leader in the system is not controlled by control protocol and the leader could not receive feedback information from the neighboring agents, which is practically impossible and each leader needs to receive the sample status of the adjacent followers.

The event-triggered control protocol in [24] is different from the above that an event-triggered time sequence of each agent is determined by the neighbor's information status in which the control protocol only calculate the event triggered threshold and the estimation error of each agent, instead of the error sums of all the neighboring agents. Using a distributed event-triggered control and low-gain feedback [27] have presented the bipartite consensus problem with input saturation for a high-order, undirected graph, and directed graph multiagent system. The author also ensured that event-triggered is employed to guarantee the semi-global bipartite consensus of the multiagent system and the lower bound of the instant time interval between the neighboring agent ensures that the Zeno behavior is neglected.

[28] provided the systematic stabilization approach for a closed-loop nonlinear system with output regulation error dynamics that is subjected to nonlinear rational with a steady-state condition and internal. The output error dynamics were eliminated via the differential-algebraic form so that a numerical optimization technique was implemented to address the controller's design parameters under bilinear matrix inequality. An event-triggered control for stabilization technique is applicable for the small-gain theorem, which demonstrates that a controlled system has a particular input-to-state stability (ISS) property. Therefore, an author [29] proposed a new event-triggered control mechanism for stabilization technique which requires so that the controlled system has achieved a certain level of input-to-state stability property, and the proposed method has further applied to design a controller for output regulation of the nonlinear system.

Robust stabilization of nonlinear system under both static and dynamic uncertainties have extensively been studied. [30] investigated a robust input-to-output stabilization for the nonlinear system has an additional external input such as perturbation. In this case, the controller was designed explicitly to specify the systems input-to-output stability (IOS) from the external input using a specified input-to-output stable (IOS) gain. Considering a global robust asymptotic stabilization problem for cascaded systems having dynamic uncertainty, [31] have developed a recursive Lyapunov design approach using induction on system relative degree so that a global stabilization will be achievable. A particular Lyapunov design of a closed-loop system leads to an iISS-Lyapunov function in a superposition form, assuming that the nonlinearities are polynomial. Through preserving the superposition form [32] proposed a changing supply function technique to remove the polynomial assumptions and generate a more general result, and the solvability conditions of the cascaded polynomials were appropriately implemented [33] using a recursive procedure.

[34] have presented a technical framework using dissipativity mechanism to solve a global output regulation problem for the nonlinear cascaded system under dynamic and static uncertainty. Besides, in the cascaded-connected system, the nonlinear robust servomechanism problem of the lower triangular system was cast into a regulation problem which led to the complete solutions of the servomechanism problem. To this respect,[35] studied the robust global stabilization of a cascaded system subjected to dynamic uncertainty utilizing a small-gain condition, in which a recursive approach was developed using the Lyapunov direct method to solve the global stabilization problem for the closed-loop system. The method has also produced an explicit Lyapunov function which is indispensable for an adaptive control scheme.

## 2.2. Output regulation

In the stabilization problem of a closed-loop system, the system's state, including the dynamic compensator, if exist, converges to the origin. Nevertheless, in some design scenarios, a weaker requirement such as 0-regulation becomes more realistic. Considering a nonlinear closed-loop system having an origin equilibrium point with zero control input and external disturbances is said to achieve 0-regulation if both the state and the output trajectory of the system converge to equilibrium point (i.e., zero), while the state of the compensator does not converge to zero. In this context, the main purpose of the regulation problem is to regulate the system output trajectory with a nonzero reference trajectory. In regulation problems, the reference trajectory could be given by the user to ensure the consensus of the single system.

It is demonstrated that a nonlinear closed-loop system with nonzero reference trajectories and control input is said to achieve regulation to the nonzero trajectory if the error between the system's output trajectory and the reference trajectory, while the system's output trajectory starting from all the initial values become zero.

So far, various literature has been studied to solve the output regulation problem, and several conclusions were also made. For instance, [18] have proposed an event-triggered and self-triggered control to research the output regulations problems using the assumption that the states of each agent are measurable. The cooperative output regulations for the linear closed-loop multiagent system are presented [36] with the actuator faults to estimate the states of the exosystem. In [37], fully-order observer-based output regulations for the closed-loop system under the switching interaction topology by dividing all agents other than external system into measurable and unmeasurable state agents, in which the controller triggering or update time and the neighbor's information transmission has to ensure the proper output regulations of the nonlinear system.

In [36], both the controller updates continuously and the neighbor communicates continuously before the output trajectory of the system converges to zero. [38] the cooperative output regulations of the closed-loop linear multiagent system using an event-triggered control strategy are introduced. In this case, an appropriate controller is designed via constructing a state observer and compensator for all agents, then the output regulations of the system are solved.

Considering the solvability and the minimum phase of the output regulations for the nonlinear system, [39,41] found a global output regulation framework for a nonlinear lower triangular system with an uncertain exosystem, and the authors have developed a framework based on non-adaptive and an expanded method and are able to establish general to regulator equations by avoiding the propagation of uncertainties.

In [40], a nonlinear robust output regulation problem-based tracking of the piezoelectric actuators with commonly existing characteristics of hysteretic. Basically, piezoelectric actuators with hysteretic characteristics are challenging to precisely control the system due to the generated unmeasurable hysteretic force, which could complicate the dynamic system. Therefore, the theoretical problem is solved by utilizing an internal model architecture and being able to asymptotically track the reference trajectory.

For the past couple of decades, the solvability of the nonlinear output regulation problem depends upon the assumption that the exosystem is linear and neutrally stable and which is the only exogenous signal accommodated by the theory of the combination of finitely many steps and sinusoidal functions. So, [42-43] proposed the global output regulation with a nonlinear exosystem to find a proper controller that can admit the signals produced by the nonlinear exosystem. Additionally, [44] has presented a global robust output regulation for the nonlinear system of the form of output feedback with an assumption that the regulator equation is polynomial. The assumption is due to the failure of finding the solutions to the complex nonlinear equations than ordinary polynomial equations. Consequently, it was illustrated that the nonlinear internal model was designed under a particular assumption which is milder than the polynomial assumption, to solve the output regulation problem.

It was depicted in [45] that the output regulations problem of the nonlinear closed system has aimed to achieve stability, asymptotic tracking, and disturbances rejection of a networked system with reference input and external disturbance. However, it has been addressed in the literature that it imposes two main challenges, such as the assumption that the solutions of the regulator are polynomial and the lack of the systematic strategy to handle the global robust output regulation problem. Understanding these challenges, the author established a general framework that could convert robust output regulation problems to the robust stabilization problem for an adequately augmented nonlinear system. Thus, using this general framework, it has been able to reduce the polynomial assumption and incorporate greater flexibility to the recent new stabilization techniques.

The output regulation problem of singular nonlinear systems using normal out feedback control was addressed in [46]. It was demonstrated here that the output regulation problem of the singular nonlinear system be solved by normal output feedback control if the system can satisfy assumptions of the normalizability and solved via singular output feedback control. This result ensures that it plays a significant role to bridges the gap between linear and nonlinear systems, and also the normal controller is also easier to implement practically.

## 2.3. Synchronization

The main intention of the cooperative control of a multiagent system is to realize coordination among agents and achieve the consensus agreement using a certain control protocol. To that respect, the synchronization problem is a typical terminology of steady-state performance that an output signal of a closed-loop system of all agents with consistent dimension synchronizes to a common trajectory. Output synchronization of a nonlinear closed-loop system is to find the decentralized control input of each agent via cooperative control protocol among the networked system so that the output of all the networked agents synchronizes an agreed trajectory. So, the nonlinear closed-loop system value is said to achieve an output synchronization, with all the agent's state starting from the initial values if the error between output trajectories of all agents and the reference trajectory as time goes by converging to zero. [47-48] proposed output synchronization problem for discrete-time nonlinear heterogeneous multiagent systems with input passivity. Then, using the proposed distributed output synchronization protocol and passivity-based control protocol, the multiagent system could be driven to synchronization, which indicates that the coupling strength between agents is weak enough. So, an event-triggered distributed synchronization protocol is employed to reduce the communication burden and update the controller's frequency.

Furthermore, [49] provided output synchronization of nonlinear dynamical networks with identically incremented-dissipative nodes. Here, an event-triggered control protocol is used to decide the communication time instants of each agent so that the system can achieve output synchronization. The network whose agent could receive global information achieves global asymptotic synchronization, and those that could only receive local information achieve bounded synchronization via applying an event-triggered controller. An adaptive neural network (NN) based output feedback synchronization controller [50] for nonlinear stochastic system driven via Levy noises with the compensated Poison and Wiener process. The presented controller ensured that all the output signals of the closed-loop systems were bounded in probability.

[51] proposed a fully distributed observer-based adaptive fault-tolerant output synchronization problem for the nonlinear multiagent system to overcome the detected unknown faults and unmeasurable states of the system. The synchronization problem can achieve fully distributed network communication for both triggering conditions and controller updates, excluding Zeno-behavior. The fixed /non-fixed time synchronization problem and event-triggered control were presented [52-54] using a combinational measurement approach. Finally, [55-57] model-based control is adopted to estimate the states of agents considering synchronization problems with intermitted network communication.

It has been noted that the synchronization of a multiagent system is by far the most twofold than that of a single agent. The autonomous synchronization of the nonlinear heterogeneous multiagent system was introduced by [58], in which the agent dynamics are neither synchronized agent dynamics nor the synchronized agent state, which are specified as priori; instead, it is determined by the initial state and inherent properties of an agent, give rise to an efficient synchronization and high degrees of adaptability. An output synchronization of the heterogeneous networked linear multiple-input-multiple-output of the multiagent system using output feedback was given [59], and [60] investigates the output synchronization of the nonlinear heterogeneous networked system via employing a novel feedforward design which relies only on the reference trajectory of the sufficiently high relative order. In this sense, the reference model for each agent was designed under proper control protocol, and the consensus was achieved. Each agent's local regulation controllers are constructed to drive each agent to the desired output trajectories, and local regulation is achieved separately. [61,65] have studied the output synchronization of nonlinear heterogeneous multiagent systems with time-varying switching communication topology by considering a scenario where the interaction graph has no spanning tree.

The studies of the output synchronization problem of the nonlinear heterogeneous multiagent system are given in [62] with the distributed sampled-data controllers. In the literature, a group of reference models was first designed to reduce the output synchronization problem of each agent into a perturbed output regulation problem. Actually, the perturbed output regulation is intended to achieve reference model tracking and disturbance rejection as given in [63] with system uncertainties. Furthermore, a general framework for the robust output synchronization of the nonlinear heterogeneous dynamics was introduced in [64]. In this respect, the agent is regarded as having heterogeneous, uncertain, and nonlinear dynamics, and each agent is collaboratively synchronized to reference trajectory, which is unknown to any agents in advance.

Reference [66] has proposed a decentralized control protocol for a group of nonlinear heterogeneous agents, providing that the synchronization for diverse motion patterns will be achieved. For the agents under investigation, with the aid of the deliberately designed reference models, the synchronization problem will be transformed into an output regulation problem. Then to ensure the validity of the proposed decentralized algorithms, the regulation problem was solved using a nonlinear output feedback control mechanism.

## 3 Conclusion

Throughout this paper, we have presented an overview of current research progress on the event-triggered control strategy for nonlinear multiagent systems via steady-state performance. Many literatures have studied the event-triggered control mechanism to overcome a frequent sampling caused by time-triggered control schemes. That means, in event-triggered control mechanisms, the agents are modulated after the system meets certain threshold conditions given in the control system. In other words, each agent's control protocol will be updated after each agent's state approach to a specific measurement value or condition; in that case, it will be guaranteed that the stable system performance and efficiency of energy utilization will be realized. An event-triggered control protocol is employed to reduce the network communications among the neighboring agent and update the controller by reducing the frequent triggering.  In the networked system, considering an event-triggered scheme, the event-generation function relies on the difference between the current state and the latest state of the system. However, some difficulties are imposed using an event-triggered control mechanism, such as abandoning many indispensable executions to overcome the continuous triggering and meet the reasonable resource allocations. The typical phenomenon of steady-state performances was divided into stabilization, regulation, and synchronization problems to discuss the stability of a closed-loop system and achieve a certain level of performance depending upon a given condition or design parameters. Various nonlinear singular and multiagent systems with heterogenous dynamics have also been investigated using the primary concept of stabilization, output regulation, and synchronization problems.